# Investigating Persuasive Socially Assistive Robot Behavior Strategies for Sustained Engagement in Long-Term Care

Cristina Getson, *Student Member, IEEE*, Goldie Nejat, *Member, IEEE*

*Abstract*— Socially assistive robots are increasingly being used to support the social, cognitive, and physical well-being of those who provide care (healthcare professionals) and those in need of care (older adults). However, the effectiveness of persuasive socially assistive robot behaviors and their impact on the sustained motivation of older adults is still not well understood. This extended abstract describes our prior human-robot interaction study on investigating the effectiveness of persuasive social robot behaviors with care providers, followed by our current research assessing the impact of these persuasive robot behaviors on the well-being of older adults in long-term care. The findings provide insights into engagement and sustained motivation of older adults when providing assistance.

## I. Introduction

Engaging in cognitive activities may help slow down cognitive decline as people age [1], impacting their well-being. However, motivation and adherence to cognitive activities for a sufficient amount of time is challenging, and there are not enough caregivers to ensure this is done on a daily basis for the long term, causing users with mild cognitive impairments (MCI) to disengage [2].

In this extended abstract, we discuss our research which explores the effectiveness of persuasive socially assistive robot (SAR) behavior strategies, specifically investigating the types of persuasive SAR behaviors that are effective at motivating and engaging older adults in a cognitive activity, sustained over the long-term. Future considerations, such as introducing large language models (LLMs) to adapt and personalize human-robot interactions to support older adults' well-being, are also discussed.

## II. Background and Research Area

### A. Persuasive Robotics in HRI

To date, persuasive strategies based on Cialdini's principles of persuasion [3] that have been used in human-robot interaction (HRI) to investigate a person's compliance to a SAR in performing a general task include: i) praise [4], ii) compliance gaining behaviors, such as emotional appeal or logic [5], iii) reciprocity [6], and iv) authority and group membership [7].

It has been found that persuasive effects in HRI can increase likeability and potentially influence compliance when SARs use praise, emotional appeal over logic, reciprocity, and group membership [4], [5], [6], [7]. Furthermore, effects are stronger when both gestures and gaze are used by the robot [8]. Authority in HRI is shown to be more persuasive when the robot is in a peer rather than authoritative role [9]. Low-controlling concrete messages enhance persuasion, while high-controlling (threatening) language decreases compliance [10].

### B. Persuasive Robotics for Healthcare

Persuasive robots have been used to investigate the effects of praise [11], [12], and expertise and goodwill [13] on compliance to perform a health-related task. However, the majority of research on persuasive SARs has evaluated persuasion factors that are not related to any particular real-world application or context [14]. Limited research has been conducted on persuasive robotics for healthcare-related tasks for older adults. There are also contradicting results on whether human persuasion theories can be successfully applied in HRI [14], in particular, for assistive HRI with vulnerable populations. Applying the wrong persuasive strategies could lead to non-compliance or adverse user behavior in older adults with MCI [10].

## III. Prior work

Our work to date addresses the overarching question: *What SAR persuasive strategies are effective at supporting sustained engagement and compliance in a cognitive activity for older adults?* Results from a mixed methods online study with care providers indicate that a combination of behavior strategies may lead to more effective engagement and compliance; these include the behavior strategies of Commitment with Emotion, Praise with Emotion, and Compliment with Anticipation [15]. Amongst the different robot behavior features displayed on the Pepper robot (body language and gestures, speech quality, eye color, and speech content), robot non-verbal behaviors received the most comments, confirming that robot body language and speech quality play an influential role in persuasive HRI [15].

## IV. Current work

We are currently investigating the long-term, sustained motivation of older adults engaged in a cognitive activity with a social robot in a long-term care environment. Specifically, using participatory design and a mixed methods approach, we will investigate how user well-being, such as enjoyment, engagement, and trust, changes over time through sustained interaction with a conversational social robot that exhibits persuasive strategies designed to motivate and engage the user in a cognitive activity day after day.

*Research supported by NSERC CREATE Healthcare Robotics (HeRo) Fellowship, Ontario Graduate Scholarship, AGE-WELL Inc., and the Canada Research Chairs Program.

C. Getson and G. Nejat are with the Autonomous Systems and Biomechatronics Lab (ASBLab), Department of Mechanical and Industrial Engineering, University of Toronto, Toronto, ON, Canada (corresponding author e-mail: cristina.getson@ mail.utoronto.ca).

## V. Future Considerations

From a conversational point of view, the introduction of LLMs for interactions can potentially increase people's well-being. For example, in [16], Woebot, a mental-health support chatbot, used a LLM incorporated into an existing rules-based system. An exploratory study comparing the LLM-enabled chatbot to one without LLMs showed that adding the LLM has the potential to make interactions more empathetic and personalized. In [17], the design of a social robot conversation system with a focus on supporting the well-being of older adults was investigated. The QT social robot used a fine-tuned GPT-3 trained to be empathetic and to ask follow-up questions to engage in conversations. The majority of participants enjoyed the interaction, were comfortable with the robot, and thought the robot was friendly. Areas of improvement for HRI include enhancing conversation flow, detecting emotions and nonverbal cues, and responding accordingly and with proper timing [17].

The cognitive abilities of older adults play an important role in social robot adaptability and personalization, which is also context-dependent. The integration of LLMs into a conversational robot could provide the needed range of social and emotional support for older adults [18]. For example, for those with moderate cognitive impairments, repeated conversations with a social robot could be helpful, including targeted instructions and pointing gestures [19]. For those with mild cognitive impairments, the robot should adapt as user needs and social situations vary, such as through *active or passive listening* [18]. For example, a user may indicate they need help to complete a task and require repeated instructions; while the user completes a task, the robot could *acknowledge* a participant's progress, and provide a variety of instructions [20] to sustain engagement. For those with no cognitive impairments, previous studies have indicated that participants think a social robot would be more helpful for others [21], such as people who live alone [22].

Finally, *privacy preservation*, *information credibility and recency* are also important factors when incorporating LLMs into social robots [18], [20], especially for vulnerable populations and those residing in long-term care homes, where the robot could be shared among multiple people.